\RequirePackage{fix-cm}
\documentclass[smallextended]{svjour3}       
\smartqed  

\usepackage{amssymb}
\usepackage{subfig}
\usepackage{graphicx}
\usepackage{color}
\usepackage{hyperref}
\usepackage{listings}
\usepackage{graphicx}
\usepackage{amsfonts}
\usepackage{braket}
\usepackage{qtree}
\usepackage{xypic}
\usepackage{dsfont}
\usepackage{mathtools}
\usepackage{lipsum, babel}
\usepackage{cite}
\begin{document}
\emergencystretch 3em
\addtolength{\jot}{3mm}

\title{A Comparison of Various Classical Optimizers for a Variational Quantum Linear Solver
}

\titlerunning{A Comparison of Classical Optimizers for a VQLS}        
\authorrunning{Pellow-Jarman et.\ al.}
\author{Aidan Pellow-Jarman \and  Ilya Sinayskiy \and  Anban  Pillay \and Francesco  Petruccione}


\institute{A. Pellow-Jarman \at
              University of Kwa-Zulu Natal, University Road, Chiltern Hill, Westville, 3629 \\
              Tel.: +27 84 628 0070\\
              \email{aidanpellow@gmail.com}\\           
              \and
          I. Sinayskiy \at
              University of Kwa-Zulu Natal, University Road, Chiltern Hill, Westville, 3629 \\
               National Institute for Theoretical Physics (NITheP), KwaZulu-Natal, 4001, South Africa\\
              \email{sinayskiy@ukzn.ac.za}\\
              \and
          A. Pillay \at
              University of Kwa-Zulu Natal, University Road, Chiltern Hill, Westville, 3629 \\
              Centre for Artificial Intelligence Research (CAIR), cair.org.za\\
              \email{pillayw4@ukzn.ac.za}\\  
              \and
          F. Petruccione \at
              University of Kwa-Zulu Natal, University Road, Chiltern Hill, Westville, 3629 \\
               National Institute for Theoretical Physics (NITheP), KwaZulu-Natal, 4001, South Africa\\
              School of Electrical Engineering, KAIST, Daejeon, 34141, Republic of Korea\\ 
              \email{petruccione@ukzn.ac.za}\\              
}

\date{Received: date / Accepted: date}

\maketitle

\begin{abstract}
Variational Hybrid Quantum Classical Algorithms (VHQCAs) are a class of quantum algorithms intended to  run on noisy intermediate-scale quantum (NISQ) devices. These algorithms employ a parameterized quantum circuit (ansatz) and a quantum-classical feedback loop. A classical device is used to optimize the parameters in order to minimize a cost function that can be computed far more efficiently on a quantum device. The cost function is constructed such that finding the ansatz parameters that minimize its value, solves some problem of interest. We focus specifically on the Variational Quantum Linear Solver (VQLS), and examine the effect of several gradient-free and gradient-based classical optimizers on performance. We focus on both the average rate of convergence of the classical optimizers studied, as well as the distribution of their average termination cost values, and how these are affected by noise. Our work demonstrates that realistic noise levels on NISQ devices present a challenge to the optimization process. All classical optimizers appear to be very negatively affected by the presence of realistic noise. If noise levels are significantly improved, there may be a good reason for preferring gradient-based methods in the future, which performed better than the gradient-free methods with only shot-noise present. The gradient-free optimizers, Simultaneous Perturbation Stochastic Approximation (SPSA) and Powell's method, and the gradient-based optimizers, AMSGrad and BFGS performed the best in the noisy simulation, and appear to be less affected by noise than the rest of the methods. SPSA appears to be the best performing method. COBYLA, Nelder-Mead and Conjugate-Gradient methods appear to be the most heavily affected by noise, with even slight noise levels significantly impacting their performance. 
\end{abstract}

\section{Introduction}
\label{intro}
Since the advent of quantum computing, many promising quantum algorithms have been developed, offering up to  exponential speed-up over their classical analogues. These include Shor's Algorithm\cite{shor} for factoring integers, the Quantum Fourier Transform\cite{qft} for the Discrete Fourier Transform, and the Harrow-Hassidim-Lloyd algorithm\cite{hhl} for solving the quantum linear systems problem. While these algorithms demonstrate the  potential of quantum computing on fault-tolerant devices, they are not yet realisable in practice, as current quantum devices are too noisy and have  too few qubits available. 

Variational Hybrid Quantum Classical Algorithms (VHQCAs) were proposed, initially with the Variational Quantum Eigensolver\cite{vqe}, as quantum algorithms that may be successfully implemented on noisy intermediate-scale quantum\cite{nisq} (NISQ) devices. VHCQAs employ a parameterized quantum circuit (ansatz) that computes a cost function, as part of a quantum-classical feedback loop, which optimizes the selection of the ansatz parameters. The cost function is constructed such that finding the ansatz parameters which minimize its value, finds the solution to some problem of interest; this corresponds to finding the ground-state of some Hamiltonian in question. The Variational Quantum Linear Solver\cite{vqls_1, vqls_2} (VQLS) is a proposed VHQCA intended to solve the quantum linear systems problem on near-term quantum devices.

The standard implementation of a VHQCA requires the selection of three components: the ansatz circuit, the cost function and the classical optimizer. The choice of classical optimizer is of interest in this paper. This choice is especially important for NISQ devices with high noise levels and shallow ansatz circuits with a limited overlap with the corresponding Hilbert space, which makes the optimization very difficult.

 Several other papers have done various different comparisons of the performance of classical optimizers in VHQCAs \cite{comparison_1, comparison_2, comparison_3, comparison_4, comparison_5}. 2 new surrogate model-based optimization methods (Model Gradient Descent and Model Policy Gradient) are introduced in \cite{comparison_1} and compared with Nelder-Mead, Bounded Optimization By Quadratic Approximation (BOBYQA) and Simultaneous Perturbation Stochastic Approximation (SPSA) in the context of the VQE. This paper\cite{comparison_1} highlights the need to use relevant cost models to make fair comparisons between optimization techniques, and the need to optimize hyper-parameters of existing classical algorithms to maximize performance. \cite{comparison_2} compares Broyden–Fletcher–Goldfarb–Shanno (BFGS), ADAM and Natural Gradient Descent in the context of the VQE. In \cite{comparison_3}, a comparison is made between 4 gradient-free optimizers, Nonlinear Optimization by Mesh Adaptive Direct Search (NOMAD), Implicit Filtering (ImFil), Stable Noisy Optimization by Branch and FIT (SnobFit) and BOBYQA in the context of the VQE. These  results also highlight the need for hyper-parameter optimization. 2 new optimization techniques are introduced in \cite{comparison_4}, combining randomized perturbation gradient estimation with adaptive momentum gradient updates to create the AdamSPSA and AdamRSGF algorithms. Single qubit optimal control is used as a test of the optimization algorithms and it is shown that these new algorithms perform exceptionally well. A comparison between Limited-memory Broyden–Fletcher–Goldfarb–Shanno (L-BFGS), Constrained Optimization by Linear Approximation (COBYLA), RBFOpt, Modified Powell's and SPSA using a noise-free implementation of the VQE to solve 6 combinatorial optimization problems is presented in \cite{comparison_5}. The results here indicate global optimization methods are more successful than local optimization methods, due to their avoidance of local optima.

This work differs significantly from the above mentioned in a variety of ways:
\begin{itemize}
    \item by collectively considering the effect of noise on the optimizer, and the choice of either gradient-based vs gradient-free optimizers, 
    \item by using a practical cost comparison between optimizers based on the resources needed to perform cost function evaluations, 
    \item by the selected set of classical optimizers being compared, and 
    \item by selecting the variational quantum linear solver as the problem of interest.
\end{itemize}
The aim of the present study is to make a practical attempt to draw a meaningful realistic comparison between a large selection of classical optimizers.
The rest of the paper is structured as follows: Section \ref{vqls_section} introduces the variational quantum linear solver and the available classical optimizers and  Section \ref{comparison} describes the experimental setup of the comparison. Section \ref{res_discussion} provides the results and a discussion, and Section \ref{conclusion} concludes.

\section{Variational Quantum Linear Solver}
\label{vqls_section}
Solving a system of $N$ linear equations with $N$ unknowns, expressible as $\mathbf{A}\vec{x}=\vec{b}$, involves finding the unknown solution vector $\vec{x}$, satisfying the equation. The quantum linear systems problem \cite{qlsp} (QLSP) is the quantum analogue to solving a system of linear equations on a classical computer, defined as follows: Let $\mathbf{A}$ be an $N\times N$ hermitian matrix (however this algorithm is not limited to a hermitian matrix; the algorithm can easily be adapted to handle the non-hermitian case) and let $\vec{x}$ and $\vec{b}$ be $N$ dimensional vectors, satisfying $\mathbf{A}\vec{x}=\vec{b}$, having corresponding quantum states $\ket{x}$ and $\ket{b}$, such that 

\begin{center}
\begin{gather}
     \ket{x}  \vcentcolon = \dfrac{\sum_{i}x_{i}\ket{i}}{\lvert\lvert \sum_{i}x_{i}\ket{i} \rvert\rvert_{2}}, \\
     \ket{b} \vcentcolon =  \dfrac{\sum_{i}b_{i}\ket{i}}{\lvert\lvert \sum_{i}b_{i}\ket{i} \rvert\rvert_{2}}.
\end{gather}
\end{center}

Given access to the matrix $\mathbf{A}$ by means of an oracle and a unitary gate $\textit{U}$ such that $\textit{U}\ket{0} = \ket{b}$, output a quantum state $\ket{x'}$ such that $\lvert\lvert\ket{x}-\ket{x'}\rvert\rvert_{2}  \leq \epsilon$, where $\epsilon$ is the error-bound of the approximate solution. This QLSP is the form of the problem solved by the HHL algorithm.

The input to the Variational Quantum Linear Solver are the matrix $\mathbf{A}$ and vector $\vec{b}$. $\mathbf{A}$ is given in a slightly different manner than in the QLSP stated above. Instead of requiring oracle access to a hermitian  matrix $\mathbf{A}$, $\mathbf{A}$ is only required to be given as a linear combination of unitary matrices $\mathbf{A}_{i}$ implemented as unitary gates, such that $\mathbf{A} = \sum_{i=1}^{n}c_{i}\mathbf{A}_{i}, c_{i} \in \mathbb{C}.$ This is equivalent to requiring that $A$ be given as a linear combination of Pauli operators. The value of $\vec{b}$ is given as a unitary gate $\textit{U}$, such that $\textit{U}\ket{0} = \ket{b}$ as per the QLSP.

\begin{figure*}
  \includegraphics[width=1\textwidth]{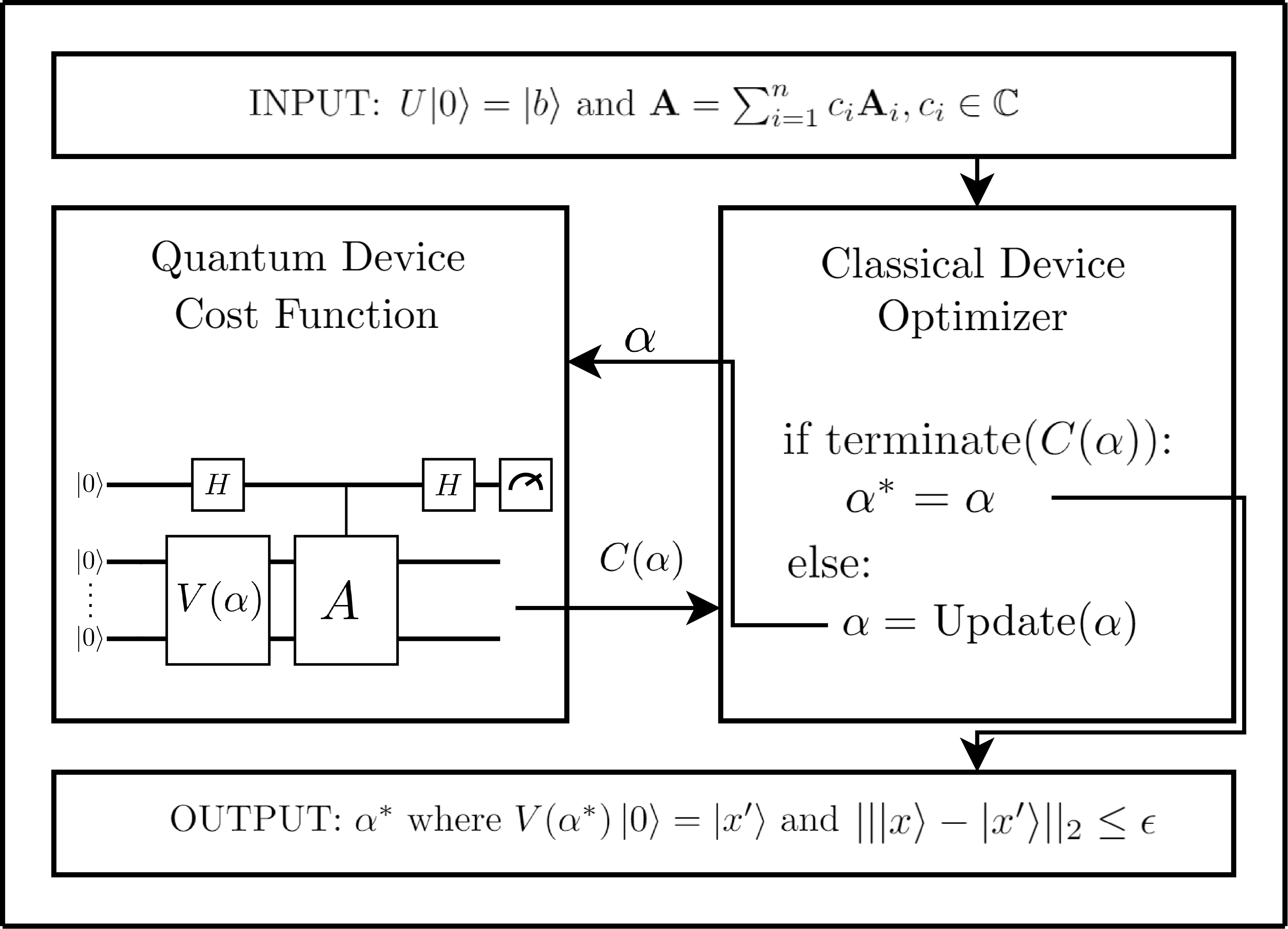}
\caption{Basic Variational Quantum Linear Solver Schematic. The algorithm runs in a simple feedback loop, whereby a classical optimizer finds the optimal parameters for the ansatz circuit, by iteratively evaluating the cost function on the quantum device and updating the parameters, until a minimum cost value is achieved. $V(\alpha)$ denotes the parameterized ansatz, $\alpha^{*}$ denotes the optimal parameters. At termination,  $V(\alpha^{*})\ket{0} = \ket{x'}$, where $\lvert\lvert\ket{x}-\ket{x'}\rvert\rvert_{2}  \leq \epsilon$, with $\ket{x}$ being the exact solution to the linear system, and $\epsilon$ being the error-bound of the approximate solution}
\label{fig:vqls_schematic}       
\end{figure*}

The VQLS algorithm runs in a simple feedback loop [Figure \ref{fig:vqls_schematic}], whereby a classical optimizer finds the optimal parameters for the ansatz circuit, by iteratively evaluating the cost function on the quantum device and updating the parameters, until a minimum cost value is achieved. Let the parameterized ansatz be denoted by $V(\alpha)$, and let the optimal parameters be denoted by $\alpha^{*}$. At the algorithms termination, $V(\alpha^{*})\ket{0} = \ket{x'}$, where $\lvert\lvert\ket{x}-\ket{x'}\rvert\rvert_{2}  \leq \epsilon$, with $\ket{x}$ being the exact solution to the linear system, and $\epsilon$ being the error-bound of the approximate solution.

The choice of classical optimizer, ansatz circuit and cost function are free to be decided in the implementation of the VQLS algorithm. The choice of classical optimizer is discussed at length and is the main focus of this paper. The ansatz choice falls into either the category of hardware-efficient (agnostic) ansatze, and problem specific ansatze. Hardware-efficient ansatze make no use of the known structure of the problem, in this case being $\mathbf{A}$ and $\vec{b}$, and instead are optimized for the specific quantum hardware available. They therefore have greater noise-resistance, making them the more realistic near-term option. Their shortfall is that there is usually no guarantee that they contain any close-approximation of the solution, due to their limited span of the relevant Hilbert space. Problem specific ansatze are designed to specifically take the structure of the problem into account, making them more suited to finding the solution to a specific problem. An example of one such ansatz is the QAOA ansatz \cite{vqls_1,vqls_2}, which involves performing repeated Hamiltonian application of Hamiltonian simulations of a \textit{driver} and \textit{mixer} Hamiltonian constructed specifically from the linear systems problem in question. These ansatze are limited by the Hamiltonian simulation which is not as near-term as the implementation of the hardware-efficient ansatz.

\subsection{Classical Optimizers}
\label{classical_opt}
There are many different classical optimizers available, both gradient-free and gradient-based, that can be used in VHQCAs for ansatz parameter optimization. Gradient-based methods require gradient information during the optimization process, which can either be obtained analytically by a quantum gradient function, or estimated by means of repeated cost function evaluations. Gradient-free methods operate as black-box optimizers, which require no extra information besides the values of the cost function evaluations. The common classical optimizers to be compared are given in Table \ref{tab:my-table}.


\begin{table}[]
\centering
\begin{tabular}{ll}

\textbf{Gradient-Free} & \\                          
Constrained Optimization by Linear Approximation$^{\alpha}$ (COBYLA) & \cite{cobyla} \\
Nelder-Mead$^{\alpha}$ & \cite{nelder-mead}  \\
Modified Powell's method$^{\alpha}$ & \cite{powell}  \\
Simultaneous Perturbation Stochastic Approximation$^{\beta}$ (SPSA)  &\cite{spsa}  \\
\\
\textbf{Gradient-Based} \\
Broyden–Fletcher–Goldfarb–Shanno algorithm$^{\alpha}$ (BFGS) &\cite{bfgs} \\
Limited-memory Broyden–Fletcher–Goldfarb–Shanno algorithm$^{\alpha}$ (L-BFGS) & \cite{l-bfgs}  \\
Conjugate Gradient method$^{\alpha}$ (CG) & \cite{cg} \\
Sequential Least Squares Programming$^{\alpha}$ (SLSQP) & \cite{slsqp} \\
ADAM$^{\beta}$ & \cite{adam} \\
AMSGrad$^{\beta}$ & \cite{amsgrad}\\
\end{tabular}
\caption{Table of Classical Optimizers Considered - The implementation of all classical optimizers are taken from the Scipy$^{(\alpha)}$ and Qiskit$^{(\beta)}$ Python libraries, under the respective functions \textbf{scipy.optimize.minimize} and \textbf{qiskit.aqua.components.optimizers}. (References given in the far-right column)}
\label{tab:my-table}
\end{table}

In this paper, the gradient-based optimizers all use an analytic gradient calculated by a quantum gradient function. It  must also be noted that the hyper-parameter settings of all classical optimizers are left as standard in their implementations, except for the `rhobeg' value of the COBYLA optimizer, which is set to $\pi$, and the learning rate of ADAM and AMSGrad, which is set to $0.1$. 

\section{\label{sec:four}Comparison Setup}
\label{comparison}
The classical optimizer comparison is explained below. The test problem insta\-nces are introduced and then the selection of the ansatz and cost function are discussed. Next the effect of noise taken into account in the simulations is explained, and finally a method for resource comparison between classical optimizers is proposed. All quantum simulations are implemented using the qiskit python module.

\subsection{Test Problem Instances}
\label{comparison_sec:1}
Here three test instances of the VQLS algorithm are presented by matrices $\mathbf{A}_{1}$, $\mathbf{A}_{2}$ and $\mathbf{A}_{3}$, of size $8\times8$, $16\times16$ and $32\times32$ respectively.
\begin{center}
$\mathbf{A}_{1} = \mathds{1} + 0.25\cdot Z_{2} + 0.15\cdot H_{3}$,

$\mathbf{A}_{2} = Z_{1} + 0.15\cdot Z_{3} + 0.5\cdot H_{4}$,

$\mathbf{A}_{3} = H_{1} + 0.25\cdot Z_{3} + 0.5\cdot X_{4}$ ,  
\end{center}
where $Z_{i}$, ($i=1,2,3$) indicates the matrix formed by the tensor product, with gate $Z$ applied to qubit $i$ and the identity gate applied to the remaining qubits. Notation is similarly defined with $H$ and $X$. $\mathds{1}$ indicates an $N \times N$ identity matrix.

In all problem instances, the state $\ket{b} = H^{\otimes n}\ket{0}^{n}$, for $n = 3, 4, 5$ respectively. The problems are real-valued, minimizing the complexity of the ansatz choice.
\subsection{Test Ansatz and Cost Function Choice}
\label{comparison_sec:2}

For each of these three test instances, a very simple hardware-efficient ansatze, consisting of parameterized $R_{y}$ single qubit rotation gates, and 2-qubit controlled-X gates, has been used. These ansatze are designed specifically to deal with real-valued problems, such as the ones under consideration. This is not a shortfall of the VQLS algorithm, and were the problem in question complex-valued, the appropriate gates could be added to the ansatz. 

\begin{figure*}
  \includegraphics[width=1\textwidth]{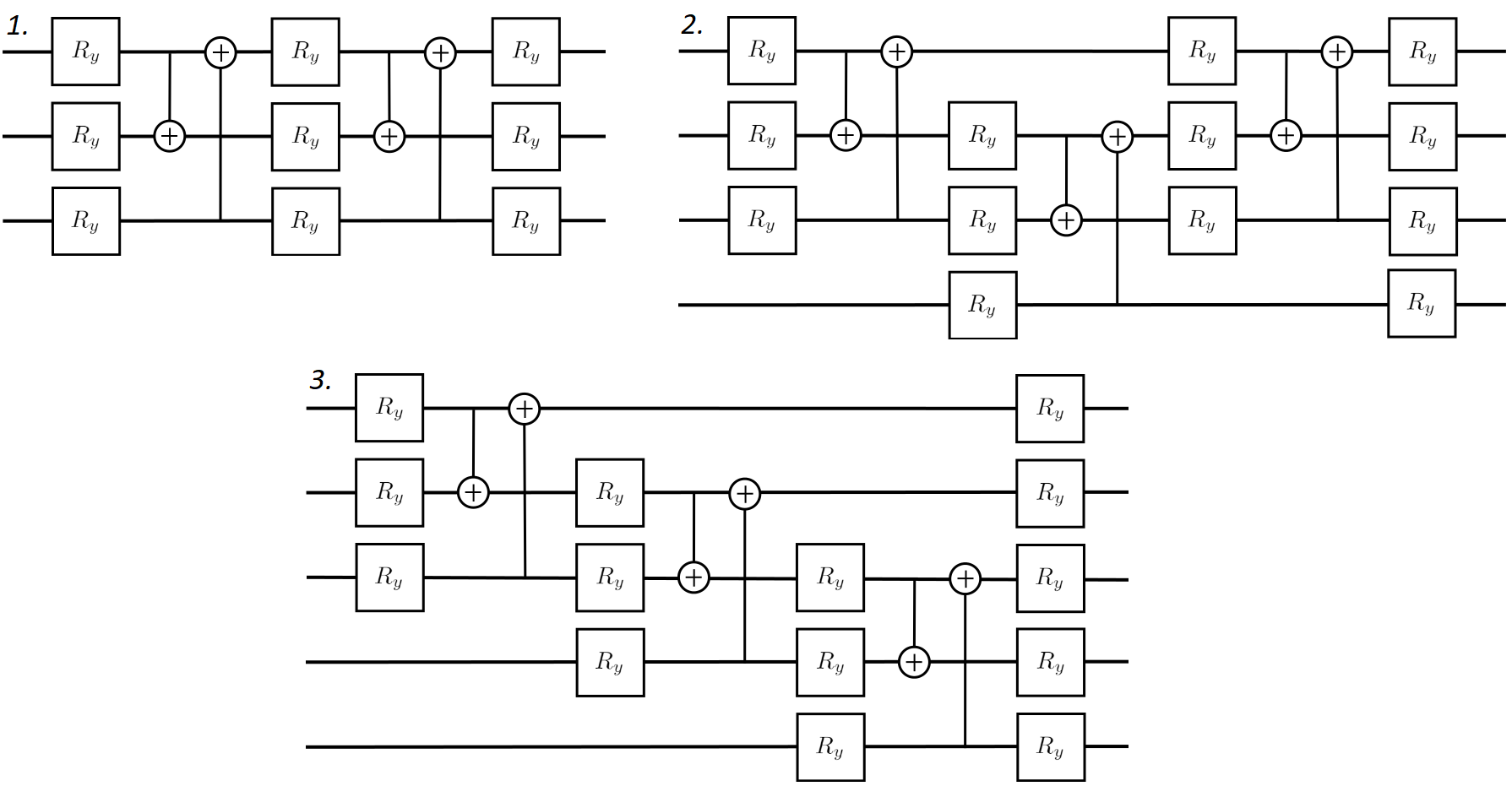}
\caption{Ansatz Circuit Diagrams: 1, 2, and 3 label the 3, 4, and 5 Qubit ansatze respectively. Each ansatz consists of parameterized $R_{y}$ single qubit rotation gates, and 2-qubit controlled-X gates}
\label{fig:ansatze}       
\end{figure*}

The ansatze are also not optimized in any way specific to these problem instances, since only the relative performance of the different classical optimizers is of interest. They are instead designed to simulate hardware-efficient ansatze, adding some level of noise resistance. They do this by making limited use of 2-qubit gates, with assumed limited backend connectivity between qubits. Here no standard backend configuration is employed; instead it is assumed only qubits between 2-qubit gates are connected.  

It is known that the initialization parameters of the ansatz can have a great impact on the optimization of its parameters, and accordingly the initial parameters have been set randomly for each of the 100 runs for the 3 problem instances, but uniformly across all of the classical optimizers and all noise levels, giving a fair comparison of the optimizers and noise levels from identical initialization parameters. Any two classical optimizers on the same problem instance, will have the same 100 random initializations used across all noise levels.

The cost function used for these VQLS instances was proposed in \cite{vqls_1} as a local version of the global cost function, intended to improve ansatz optimization by avoiding the exponentially vanishing gradients that global cost functions experience as the number of qubits increases. This local cost function is given by, 
\begin{center}
$C_{L}=\dfrac{\bra{x}H_{L}\ket{x}}{\bra{x}A^{\dagger} A\ket{x}}$,
\end{center}
with the effective Hamiltonian $H_{L}$ given by,
\begin{center}
$H_{L} = A^{\dagger}U(\mathds{1}-\frac{1}{n}\Sigma^{n}_{j=1}\ket{0_{j}}\bra{0_{j}}\otimes\mathds{1}_{\Bar{j}})U^{\dagger}A$.
\end{center}
This cost function is also selected here because it does not require controlled application of the ansatz, making it more resistant to noise. In depth analysis of this cost function as well as the computation of analytic gradients for this cost function are given by \textit{Bravo-Prieto et al.} in \cite{vqls_1}. 

\subsection{Effect of Noise}
\label{comparison_sec:3}
To demonstrate the effect of noise on classical optimizers, three different quantum simulations were used. The first was a state vector simulation, having zero noise, implemented using qiskit's \textit{Statevector Simulator}. The second simulation was a perfect fault-tolerant quantum device, using qiskit's \textit{Qasm Simulator} having only shot-noise (that is the noise occurring through the sampling of expectation values in the Hadamard Test, with number of shots being set to 10 000). The amount of shot-noise is inversely proportional to the number of shots, with the state vector simulation being equivalent to a perfect fault-tolerant quantum device with an infinite number of shots. The third simulation is that of a realistic near-term quantum device having shot-noise as well as realistic noise model sampled from the IBM Vigo quantum device, and added to qiskit's \textit{Qasm Simulator}. This simulation is intended to provide the most difficult and realistic optimization challenge for the classical optimizers.

\subsection{Resource Comparison}
\label{comparison_sec:4}
The above mentioned classical approaches were compared in terms of the number of evaluations of the cost function and, for optimizers making use of gradient function calls, the number of evaluations of the gradient function. The cost of the quantum resources required for one gradient function evaluation can be equivalently expressed in terms of a number of cost function evaluations, with one gradient function evaluation being equivalent to two evaluations of the cost function per parameter in the ansatz, as per the analytic gradient function procedure\cite{vqls_1, gradient}. Comparing classical optimizers by the number of cost function calls gives an objective way to measure the amount of quantum resources required by a classical optimizer during optimization. For the 3, 4 and 5 qubit problem instances, the optimizers are limited in this comparison to 600, 800 and 1000 evaluations of the cost function. Classical optimizers were terminated if they reach this cost function evaluation limit. 

\section{Results and Discussion}
\label{res_discussion}
\begin{figure*}[h]
\includegraphics[width=1\textwidth]{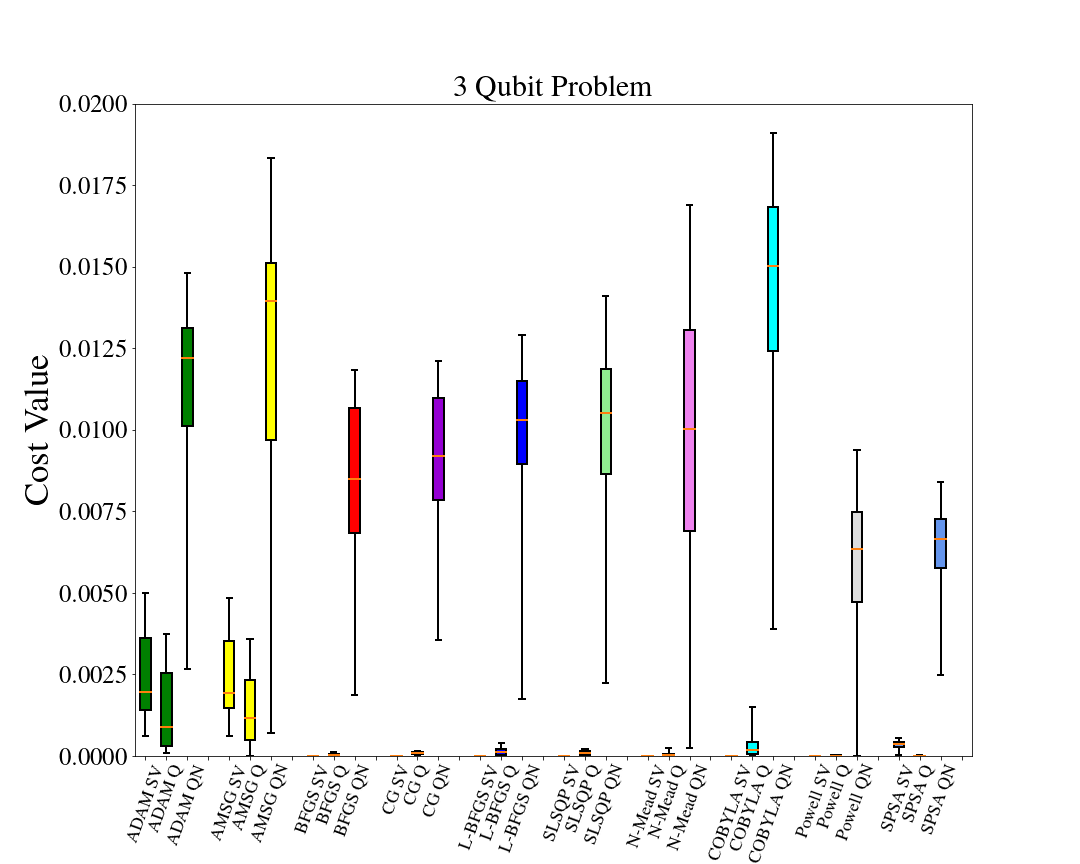}
\caption{3 Qubit Problem Instance - Termination Value Box plot: This box plot captures the final distribution of the cost function value at termination, for the 50 best runs, for all optimizers, for all noise-levels. The lower cost value is better (SV - Statevector, Q - Qasm, QN - Qasm + Noise)}
\label{fig:boxplot_3}
\end{figure*}
\begin{figure*}[h]
\includegraphics[width=1\textwidth]{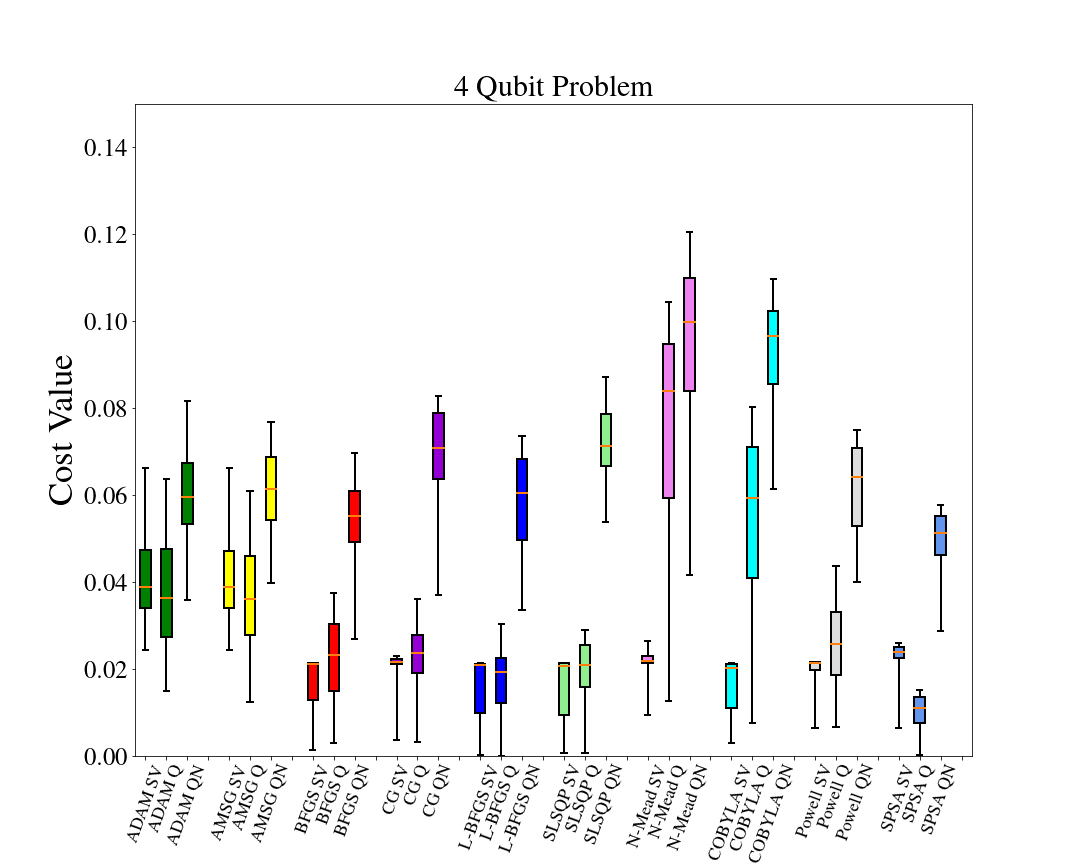}
\caption{4 Qubit Problem Instance - Termination Value Box plot: This box plot captures the final distribution of the cost function value at termination, for the 50 best runs, for all optimizers, for all noise-levels. The lower cost value is better (SV - Statevector, Q - Qasm, QN - Qasm + Noise)}
\label{fig:boxplot_4}
\end{figure*}
\begin{figure*}[h]
\includegraphics[width=1\textwidth]{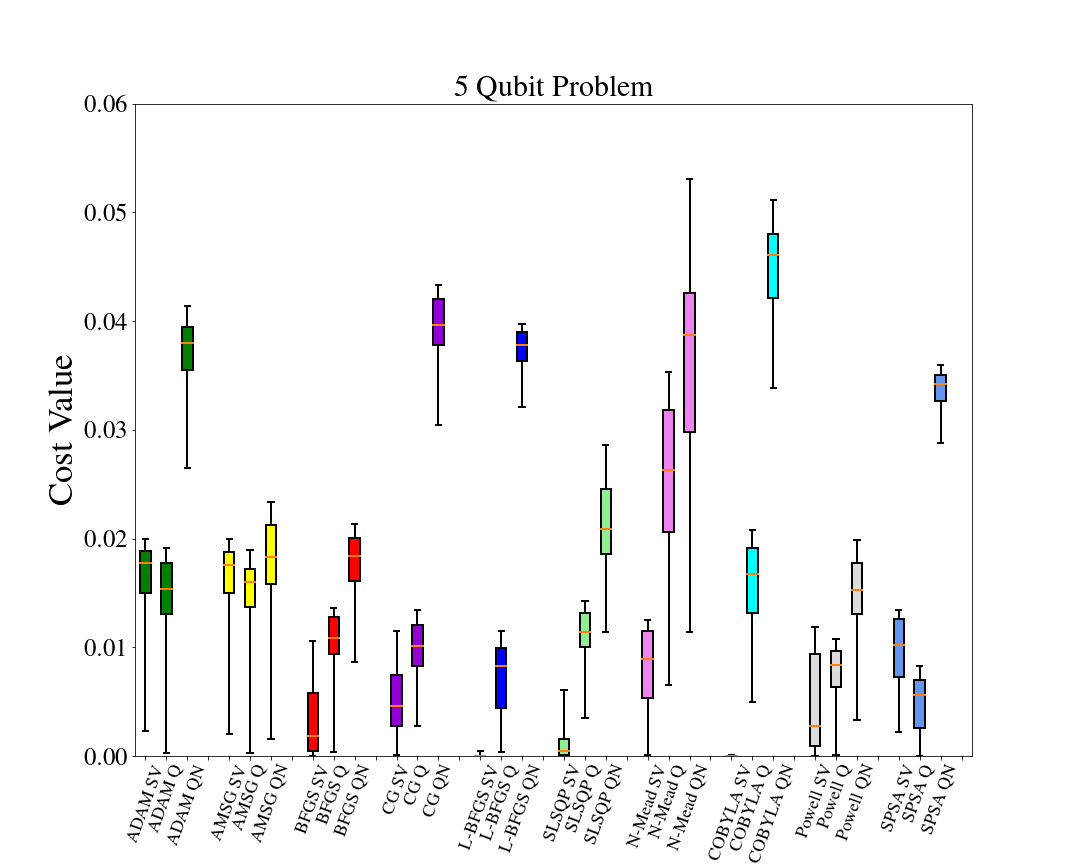}
\caption{5 Qubit Problem Instance - Termination Value Box plot: This box plot captures the final distribution of the cost function value at termination, for the 50 best runs, for all optimizers, for all noise-levels. The lower cost value is better (SV - Statevector, Q - Qasm, QN - Qasm + Noise)}
\label{fig:boxplot_5}
\end{figure*}

\begin{figure*}[htb]
\includegraphics[width=\textwidth]{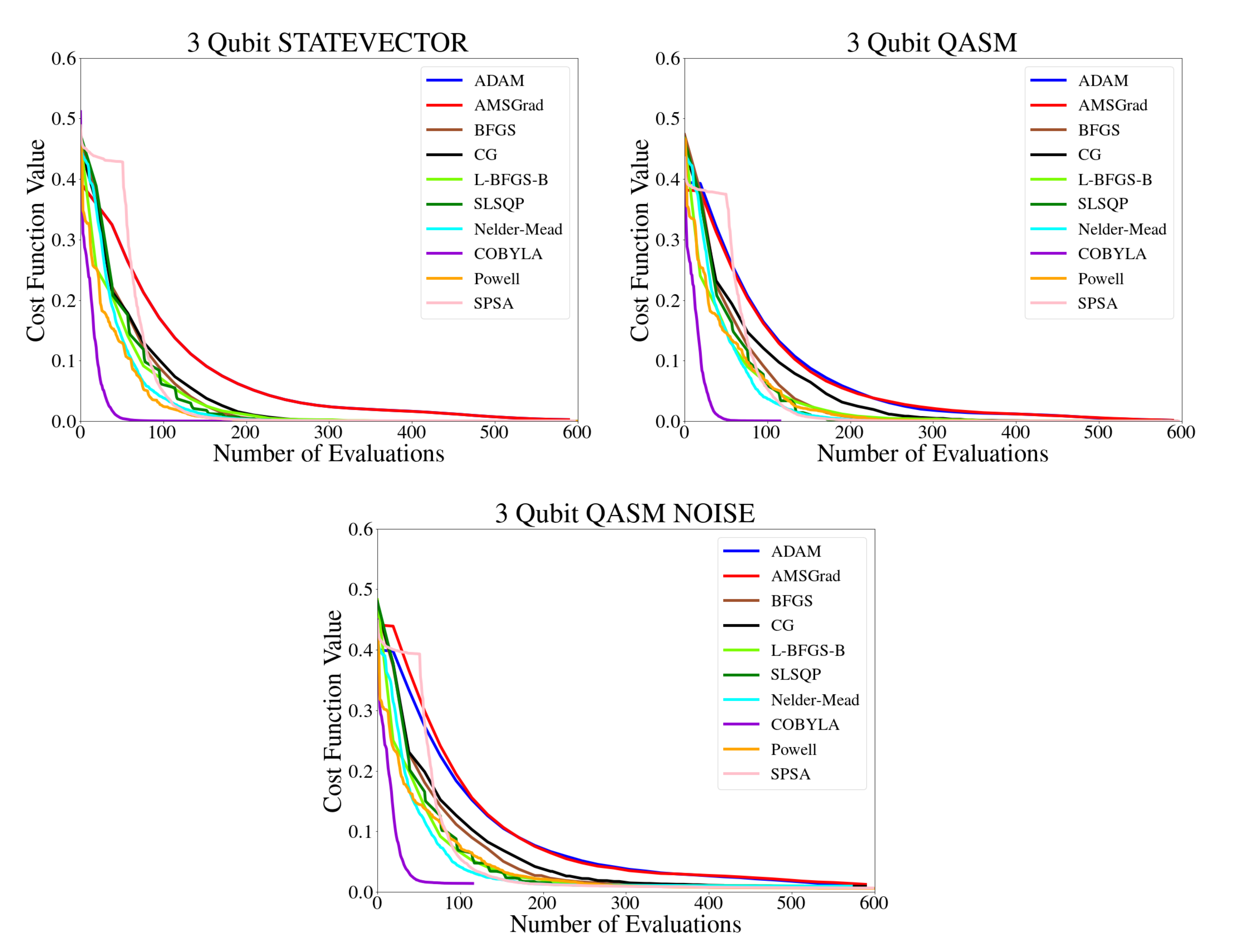}
\caption{3 Qubit Problem Instance - Convergence Rate: The average rate  of convergence across the 50 best attempts, for all classical optimizers, at each noise level}
\label{fig:line_3}
\end{figure*}

\begin{figure*}[htb]
\includegraphics[width=\textwidth]{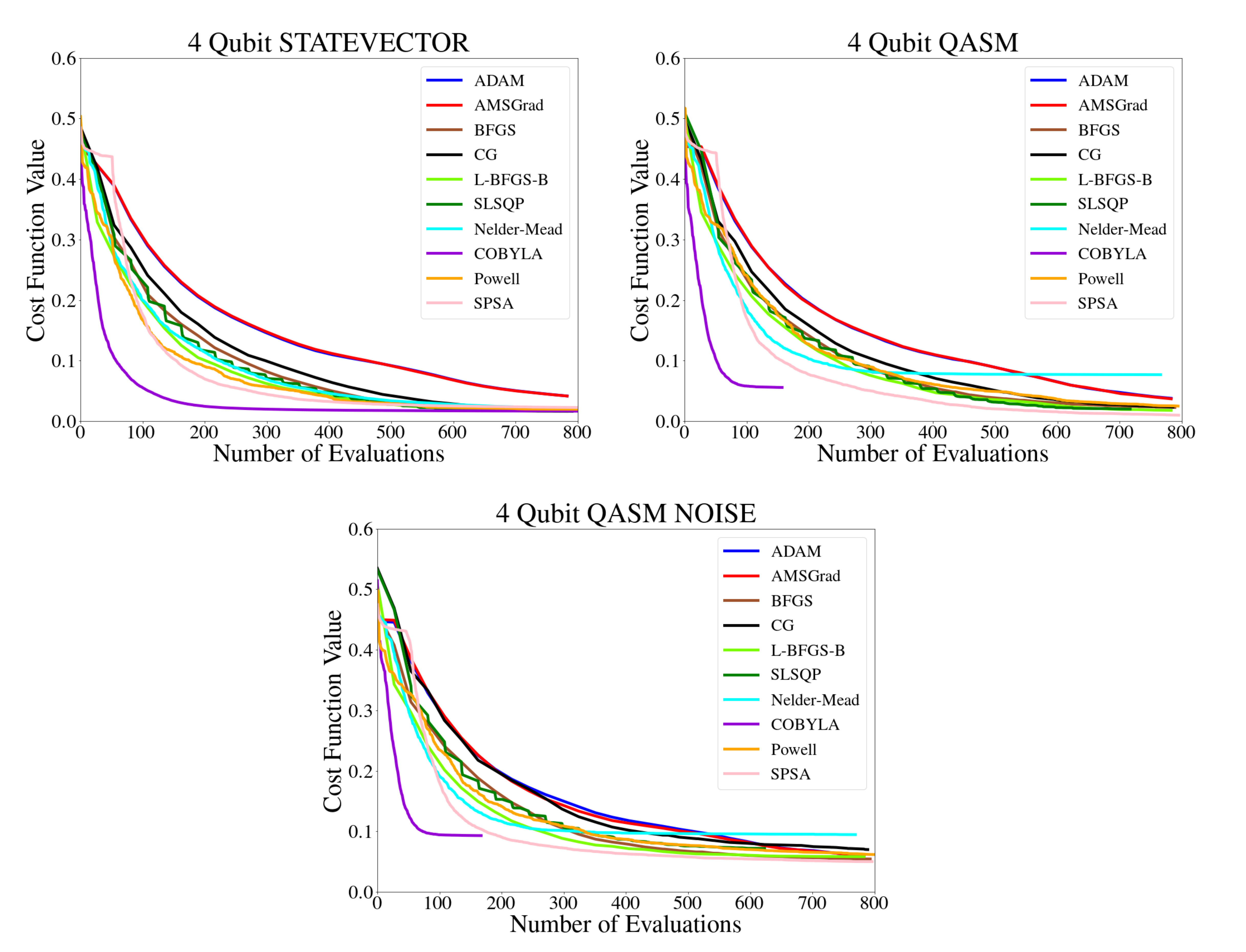}
\caption{4 Qubit Problem Instance - Convergence Rate: The average rate  of convergence across the 50 best attempts, for all classical optimizers, at each noise level}
\label{fig:line_4}
\end{figure*}

\begin{figure*}[htb]
\includegraphics[width=\textwidth]{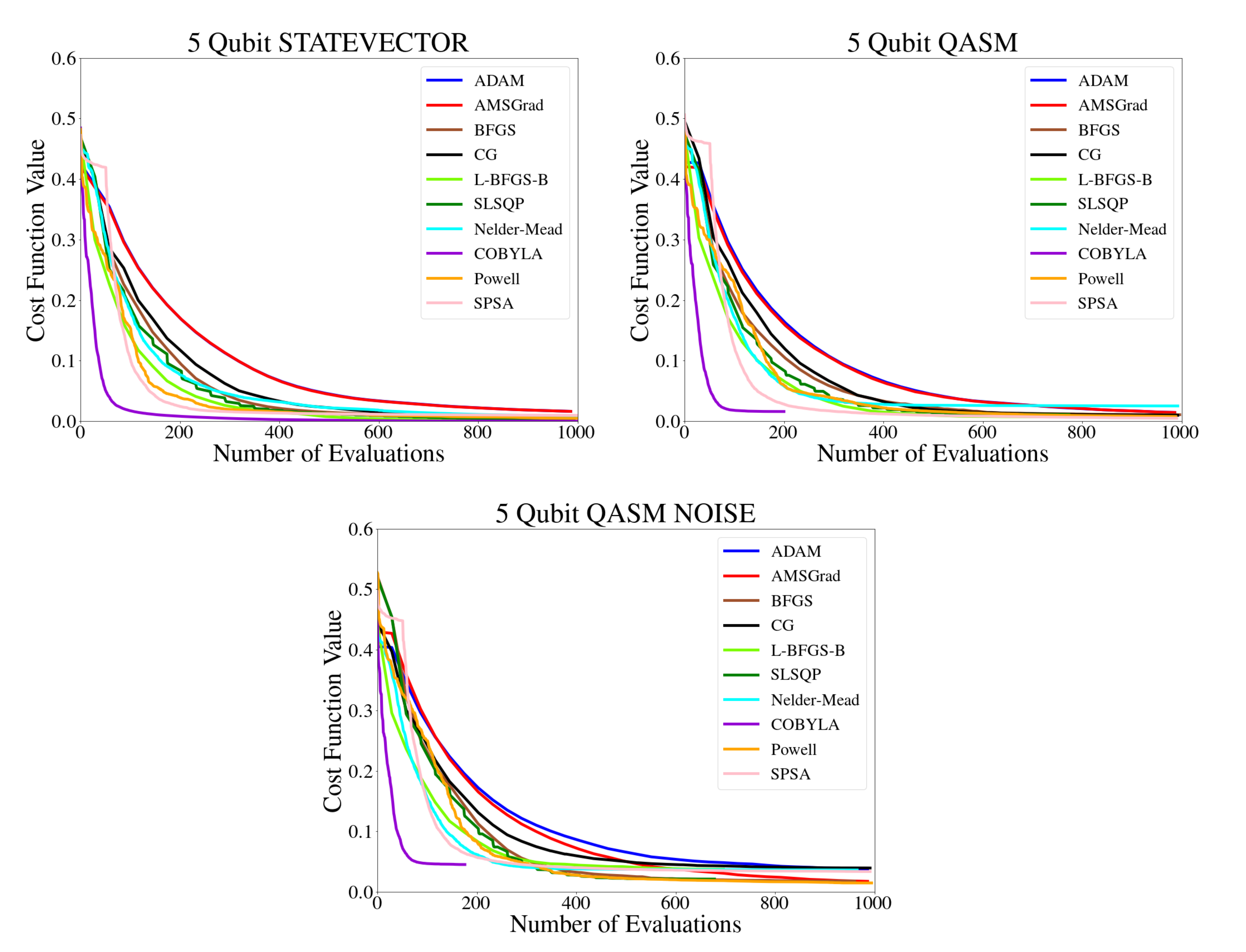}
\caption{5 Qubit Problem Instance - Convergence Rate: The average rate  of convergence across the 50 best attempts, for all classical optimizers, at each noise level}
\label{fig:line_5}
\end{figure*}

Box plots [Fig: \ref{fig:boxplot_3}, \ref{fig:boxplot_4}, \ref{fig:boxplot_5}] and line graphs [Fig: \ref{fig:line_3}, \ref{fig:line_4}, \ref{fig:line_5}] record the final distribution of the cost value and average rate of convergence respectively, for the top 50 best performing optimizations out of the 100 for each problem instance and noise level. The top 50 were used in order to exclude any of the non-converging optimization attempts that commonly occurred. These non-converging attempts failed to converge almost at all from their initial cost values. This does not present so much of an issue for optimizers in general as these non-converging optimization attempts are clearly visible after the first few iterations, and so the optimization process can just be restarted from another random initialization of the ansatz parameters. 

Near uniform trends are observed across all test problem instances with regards to the final distribution of the cost value at termination [Fig: \ref{fig:boxplot_3}, \ref{fig:boxplot_4}, \ref{fig:boxplot_5}], and therefore these results are discussed in general with respect to the trends observed in performance across problem instances as noise levels increase, from statevector simulation, to include shot-noise, and finally to include a noise model sampled from a real quantum device. 

In the statevector simulation, for all problem instances, almost all classical optimizers performed similarly well, with ADAM and AMSGrad slightly under performing, and the gradient-based BFGS, L-BFGS and SLSQP, and the gradient-free COBYLA, slightly out performing the rest. There was not much difference between the gradient-free and gradient-based approached in general in a compl\-etely noise-free simulation. 

Once shot-noise is taken into account, the gradient-based optimizers appear to outperform the gradient-free optimizers on average for all problem instances, with the exception of SPSA, which performed the best on average in all problem instances. The remaining gradient-free optimizers performance got noticeably worse in the presence of shot-noise alone. Again BFGS, L-BFGS and SLSQP performed the best for the gradient-based optimizers, which all in general were less affected by the shot noise. ADAM and AMSGrad remain the under-performers in the gradient-based optimizers.  

Finally when a realistic noise model is added to the simulation, there is a clear shift in the results. While all of the classical optimizers perform significantly worse under the increased noise levels, some optimizers handle the increased noise levels much better than others. The gradient-based optimizers may perform worse under realistic noise levels when compared to their performance under shot-noise because it may become increasingly difficult for the gradient-based optimizers to calculate analytic gradients with the increased noise levels, as calculating analytic gradients requires very many circuit evaluations (2 per ansatz parameter). Again SPSA appears to be the best performing classical optimizer overall, with Powell's method also performing compara\-tively well within the gradient-free optimizers. AMSGrad and BFGS also perform exceptionally well under the increased noise-levels within the gradient-based optimizers. COBYLA, Nelder-Mead and Conjugate-Gradient, which had both performed well in the statevector simulation, perform the worst in the noisy simulation.

Looking at the average convergence rate of the classical optimizers [Fig: \ref{fig:line_3}, \ref{fig:line_4}, \ref{fig:line_5}] there is a general trend of near equally fast convergence between all approaches, in each respective problem instance and noise level. The relative performance of the optimizers is also uniform across the problem instances. The decision to use gradient-free or gradient-based approaches appears to be of little consequence with respect to convergence speed as well. ADAM, AMSGrad and COBYLA serve as exceptions to the general trend, with ADAM and AMSGrad consistently converging the slowest, and COBYLA consistently converging the fastest. COBYLA also terminates early in the presence of any noise. 

\section{Conclusion}
\label{conclusion}
It is clear that the realistic noise levels on NISQ devices present a big challenge to the optimization process. In very low noise levels, gradient-based approaches appear to outperform gradient-free optimizers. The performance of classical optimizers under realistic noise levels varies heavily within either of the sets of gradient-free or gradient-based approaches. There is no clear winner in general between gradient-free and gradient-based approaches under realistic noise levels. SPSA appears to perform the best under realistic noise levels, with Powell's method also performing relatively well in the gradient-free optimizers, and AMSGrad and BFGS performing the best in the gradient-based optimizers. Some optimizers, notably COBYLA, Nelder-Mead and Conjugate Gradient, which performed well in the noise-free environment, perform exceptionally worse in the noisy environment.  

It must be noted that these results are simply an initial investigation into the relative performance of various classical optimizers available. Not much focus was paid to the setting of the hyper-parameters of the various algorithms, and as such it may be possible to greatly improve the performance of those optimizers which are heavily dependant on these by selecting appropriate hyper-parameter settings. This is intended for a future work. These results show the relative performance of the various algorithms under differing noise levels while keeping hyper-parameters constant. As this work only employed quantum simulators, confirming these results on real quantum hardware has also been left aside for a future work. The results of this work are based upon the performance of the various classical optimizers in a variational quantum linear solver; it is left to a future work to investigate whether similar results are observed in other variational algorithms.

\begin{acknowledgements}
This work is based upon research supported by the South African Research Chair Initiative, Grant No. UID 64812 of the Department of Science and Innovation of the Republic of South Africa and National Research Foundation of the Republic of South Africa.
\end{acknowledgements}

\clearpage 
%
%



\end{document}